\documentclass[iop]{emulateapj}
\usepackage{natbib}
\defcitealias{kempner2004}{KD04}
\defcitealias{owers2011}{O11}
\defcitealias{smith2010}{S10}

\usepackage{epsfig}
\usepackage{rotating}
\usepackage{graphics}
\newcommand{\chan}{{\it Chandra}}
\newcommand{\ha}{$\rm{H}{\alpha}$}
\newcommand{\hb}{$\rm{H}{\beta}$}

\newcommand{\kms}{$\,\rm{km\,s^{-1}}$}

\shorttitle{Shocking Tails in Abell~2744}
\shortauthors{Owers et al.}
\begin{document}
\title{Shocking Tails in the Major Merger Abell~2744.}
\author{Matt S. Owers\altaffilmark{1,2,3}, Warrick J. Couch\altaffilmark{1}, Paul E.J. Nulsen\altaffilmark{4}, Scott W. Randall\altaffilmark{4}}
\altaffiltext{1}{Center for Astrophysics and Supercomputing, Swinburne University of Technology, Hawthorn, VIC 3122, Australia; mowers@aao.gov.au}
\altaffiltext{2}{Australian Astronomical Observatory, PO Box 296, Epping, NSW 1710, Australia}
\altaffiltext{3}{Australian Research Council Super Science Fellow}
\altaffiltext{4}{Harvard Smithsonian Center for Astrophysics, 60 Garden Street, Cambridge, MA 02138, USA}

\begin{abstract}
We identify four rare ``jellyfish'' galaxies in Hubble Space Telescope imagery of the major merger cluster Abell~2744. These galaxies harbor trails of star-forming knots and filaments which have formed in-situ in gas tails stripped from the parent galaxies, indicating they are in the process of being transformed by the environment. Further evidence for rapid transformation in these galaxies comes from their optical spectra, which reveal starburst, poststarburst and AGN features. Most intriguingly, three of the jellyfish galaxies lie near ICM features associated with a merging ``Bullet-like'' subcluster and its shock front detected in \chan\ X-ray images. We suggest that the high pressure merger environment may be responsible for the star formation in the gaseous tails.
This provides observational evidence for the rapid transformation of galaxies during the violent core passage phase of a major cluster merger.
\end{abstract}


\section{Introduction}

The hierarchical nature of large-scale structure formation is spectacularly revealed by observations of approximately equal mass mergers between pairs of galaxy clusters. These major mergers can subject galaxies to an environment that can drive abnormal rates of galaxy evolution, particularly at times close to pericentric passage \citep{roettiger1996, bekki2010}. For example, it has been suggested that the core passage phase of a merger may be important in both triggering and truncating starbursts \citep{caldwell1993, caldwell1997,poggianti2004,hwang2009, ma2010}, although it remains unclear which merger-specific mechanisms are responsible. Simulations suggest that the high pressure environment a galaxy encounters during the core passage phase of a merger may result in the collapse of giant molecular clouds (GMCs) within a galaxy, leading to an initial burst of star formation \citep{bekki2003, kronberger2008, bekki2010} while the subsequent stripping of the interstellar medium terminates GMC formation, halting further star formation \citep{fujita1999}. Evidence for these processes should reveal itself in the vicinity of observed shock fronts which mark regions of severe merger activity. In this context, \citet{chung2009} find that the shock front in the ``Bullet'' cluster \citep[1ES0657-558][]{markevitch2004} has not had an appreciable effect on the star formation in the galaxies in its vicinity. Thus, while it appears that the intense core passage phase of a major merger may play a significant role in shaping the star-forming properties of the galaxies, more detailed observations are required to understand the dominant processes.

In this vein, the merging cluster Abell~2744 \citep[z=0.3064, hereafter A2744;][hereafter O11]{owers2011} is an excellent candidate for testing the effects of major mergers on cluster galaxies. Recently, \citetalias{owers2011} combined \chan\ X-ray and Anglo Australian Telescope AAOmega optical observations to constrain the dynamics of the merger in A2744. The new data allowed a refinement of the previous merger scenarios \citep{kempner2004,boschin2006} and indicate that a high velocity ($\sim  4750$\kms), near head on, major merger along a roughly north-south axis at $\sim27^\circ$ to our line of sight is shortly past core passage,  along with an infalling group to the northwest \citep[see also][]{merten2011}. The \chan\ data reveal evidence for a ``Bullet-like'' shock front driven by the remnant core of the less massive subcluster.  Its higher velocity dispersion and strong lensing features around its brightest member (Figure 1) indicate that the northern-most subcluster is the remnant of the more massive subcluster. The properties of the major merger in A2744 make it a prime target for understanding how this extreme environment affects the resident galaxies.

In this Letter we present a suggestive spatial coincidence between a merger-affected region revealed by the \chan\ analysis of \citetalias{owers2011} and several peculiar galaxies discovered in archival {\it Hubble Space Telescope} (HST) imagery. These ``jellyfish'' galaxies exhibit one-sided  trails of extremely blue knots and filaments reminiscent of those first noticed by \citet{owen2006} and \citet[][see also \citealp{sun2007,yoshida2008,yagi2010,smith2010,yoshida2012}]{cortese2007}. These knots and filaments are interpreted as the manifestation of hot, young stars formed {\it in-situ} within gas which has been stripped from the parent galaxy \citep[][hereafter S10]{smith2010}. We suggest that the major merger environment of A2744 is significantly influencing these `galaxies in turmoil'.

\section{Observations}

\begin{figure*}
{\includegraphics[angle=0,width=0.95\textwidth]{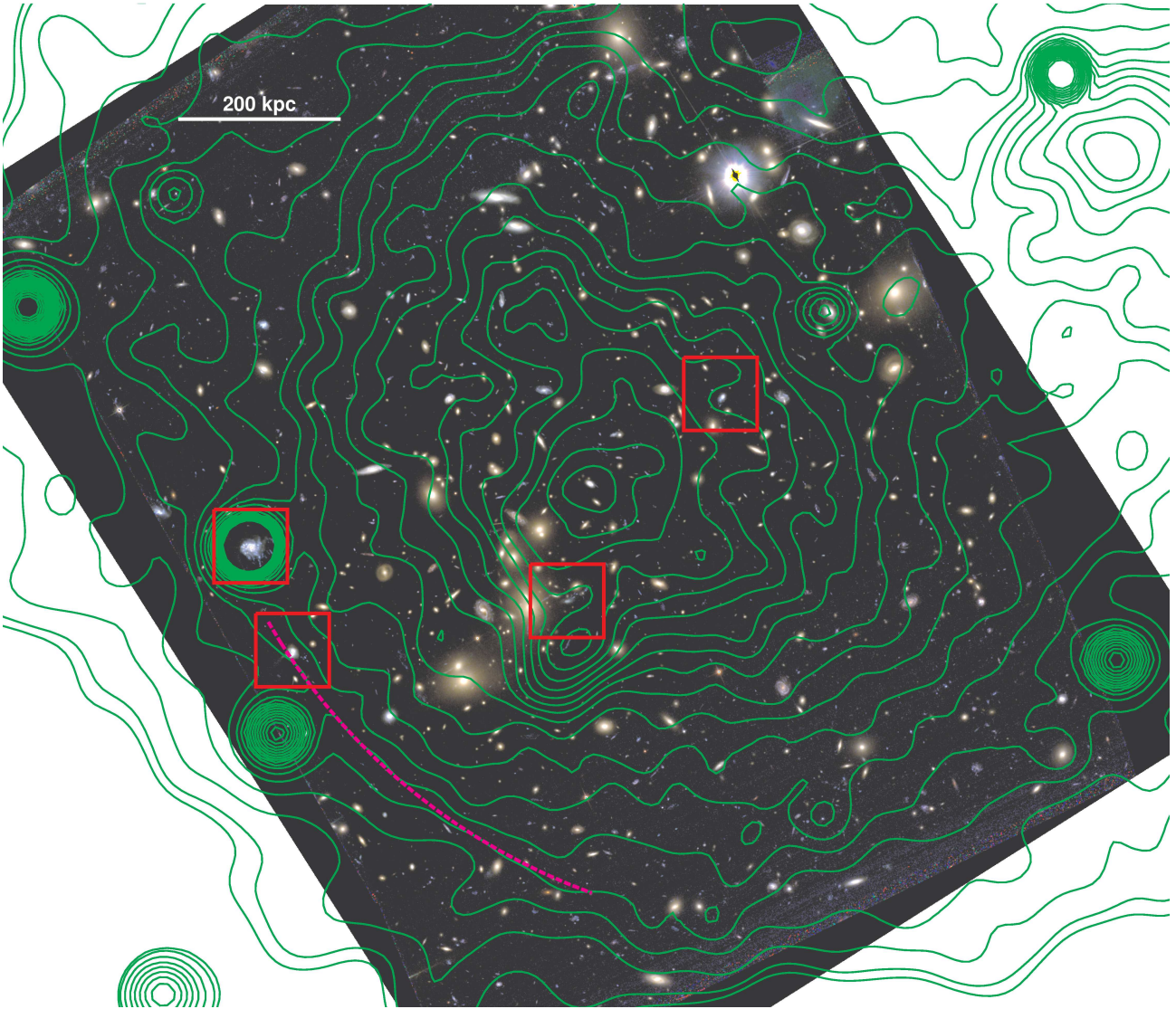}}
\caption{HST/ACS RGB image of A2744. The green contours show \chan\ X-ray surface brightness. The red boxes highlight the ``jellyfish'' galaxies which are shown in more detail in Figure~\ref{rgbimages}. From east to west, the four galaxies are F0083, F0237, the ``central'' jellyfish and F1228. The magenta dashed curve shows the shock edge reported in \citetalias{owers2011}.}
\label{hst_plus_chan}
\end{figure*}

HST/ACS observations of A2744 were taken during 2009 October using the F435W, F606W and F814W filters. The observations consist of north and south pointings each having exposure times of 6624\,s for the F606W and F814W filters and 8081\,s for the F435W filter. The data were retrieved from the archive and the pointings were combined using the {\sf multidrizzle} task \citep{koekemoer2002}. The RGB color image is shown in Figure~\ref{hst_plus_chan}, where the F814W image is used for the red channel, F606W green and F435W blue. Overlaid are contours from our background-subtracted, exposure-corrected \chan\ X-ray image \citepalias{owers2011}. Plotted as a magenta arc is the position of the shock edge described in \citepalias{owers2011}. We also make use of our AAOmega spectra.

\section{Results and Analysis}

\subsection{HST/ACS imaging}

\begin{turnpage}
\begin{figure*}
{\includegraphics[angle=0,width=0.24\paperwidth]{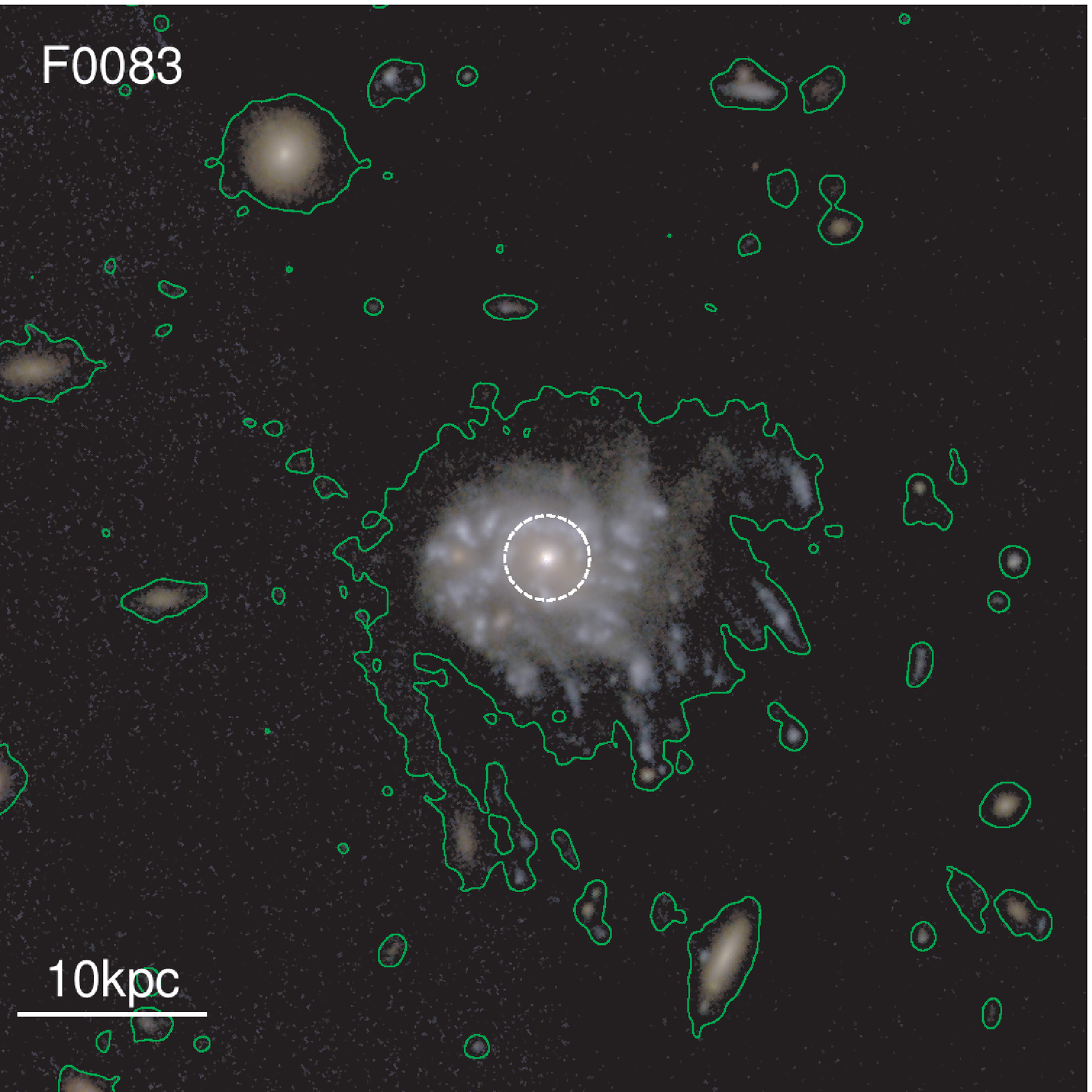}}
{\includegraphics[angle=0,width=0.24\paperwidth]{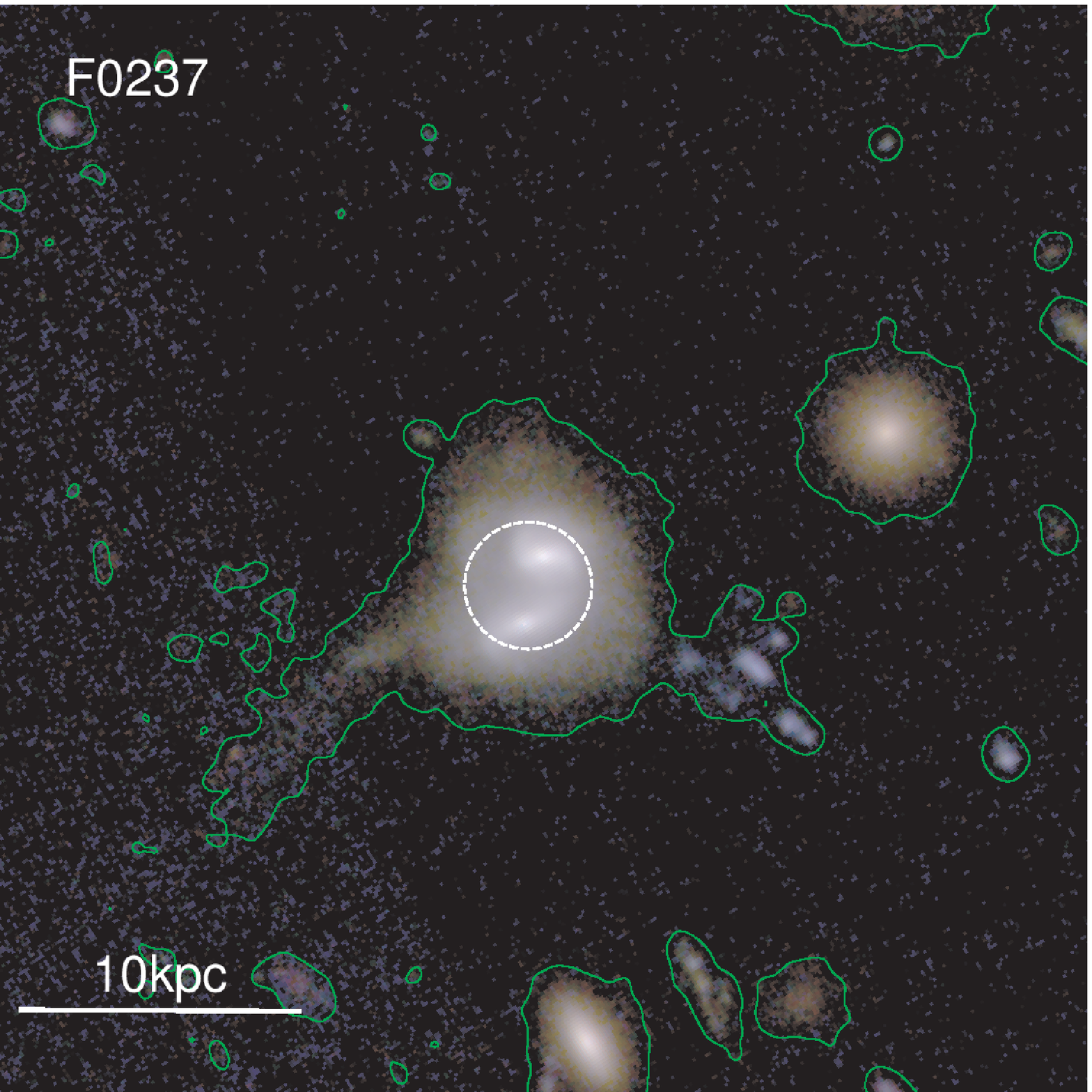}}
{\includegraphics[angle=0,width=0.24\paperwidth]{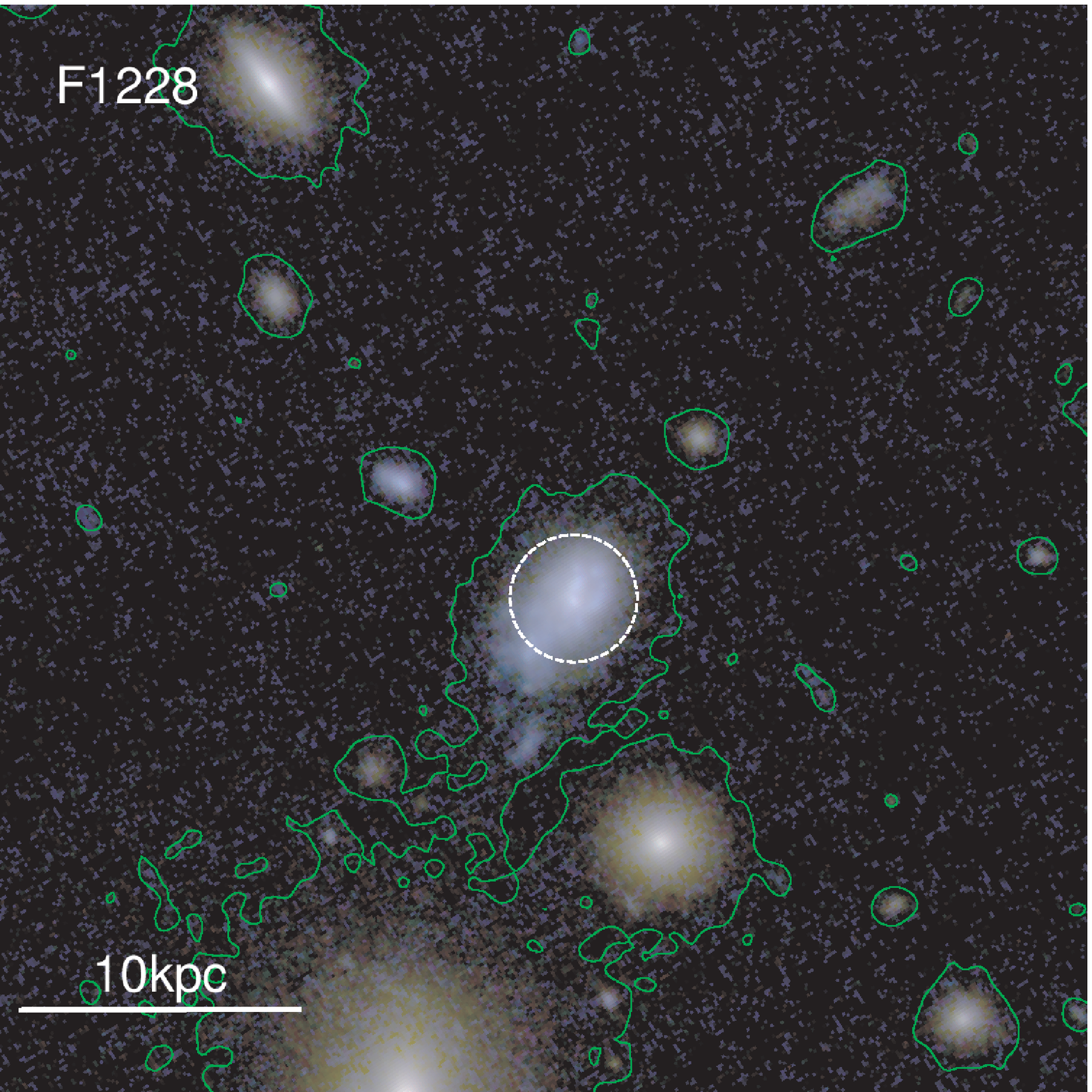}}
{\includegraphics[angle=0,width=0.24\paperwidth]{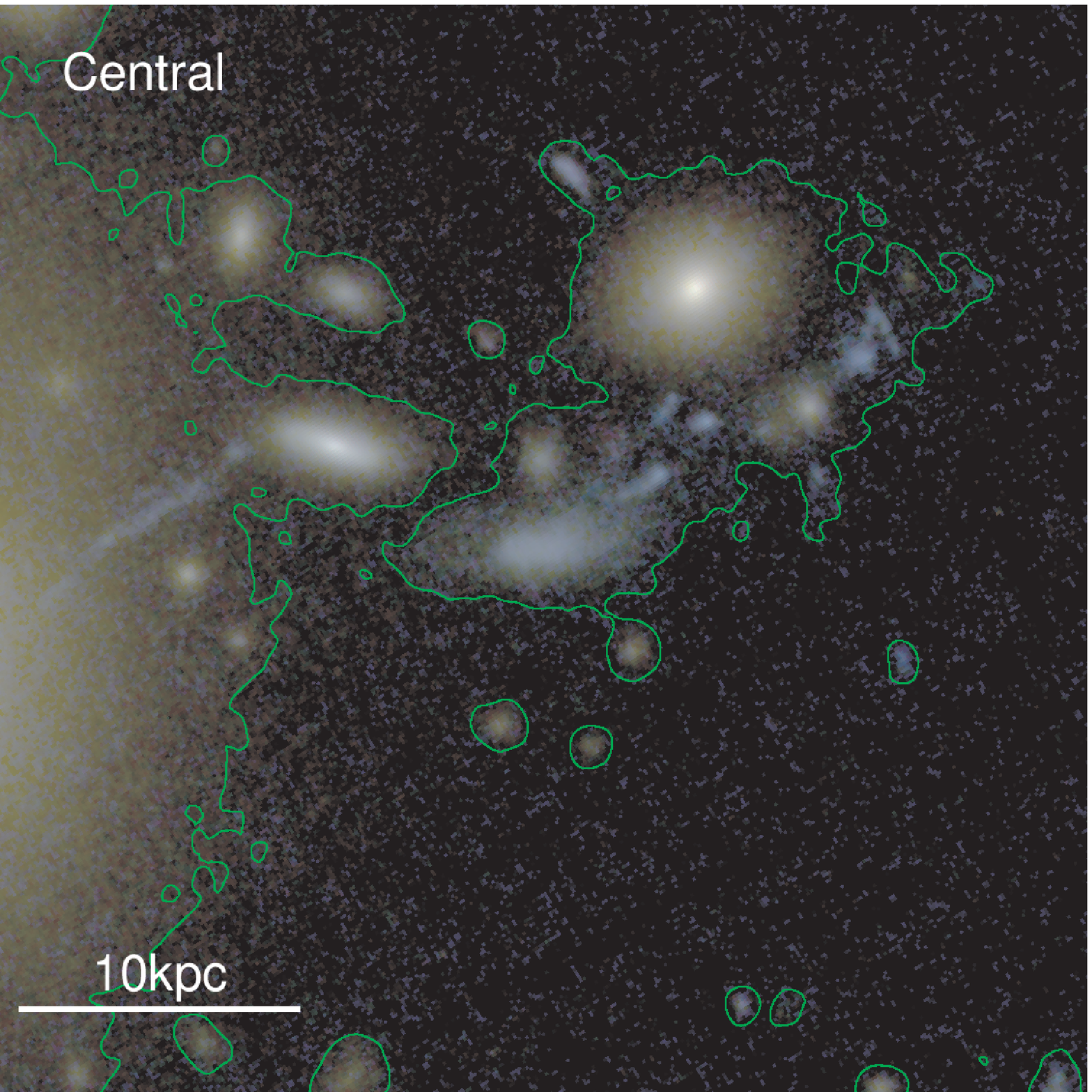}}\\
{\includegraphics[angle=90,width=0.32\paperwidth,height=0.25\paperwidth]{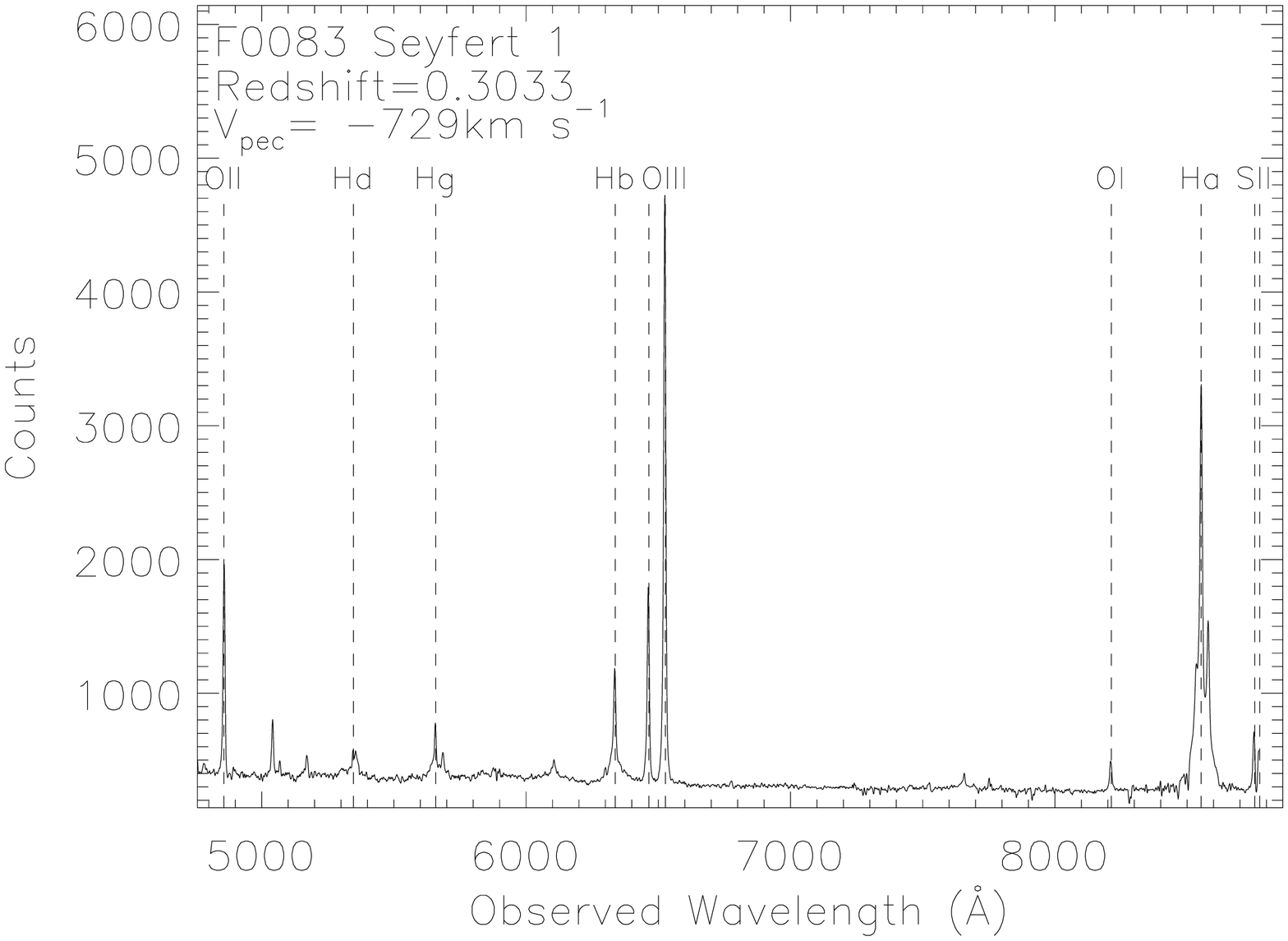}}
{\includegraphics[angle=90,width=0.32\paperwidth,height=0.25\paperwidth]{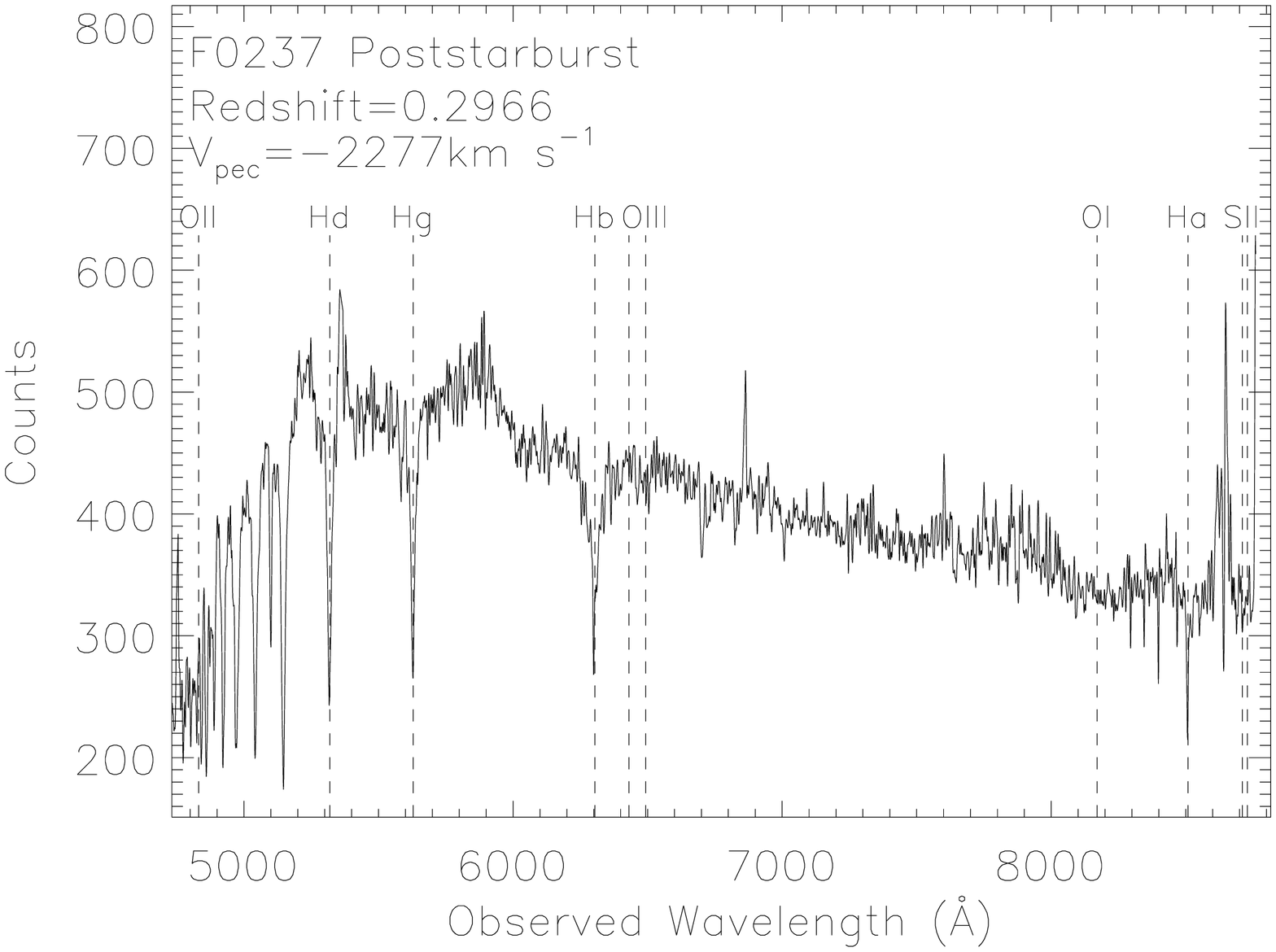}}
{\includegraphics[angle=90,width=0.32\paperwidth,height=0.25\paperwidth]{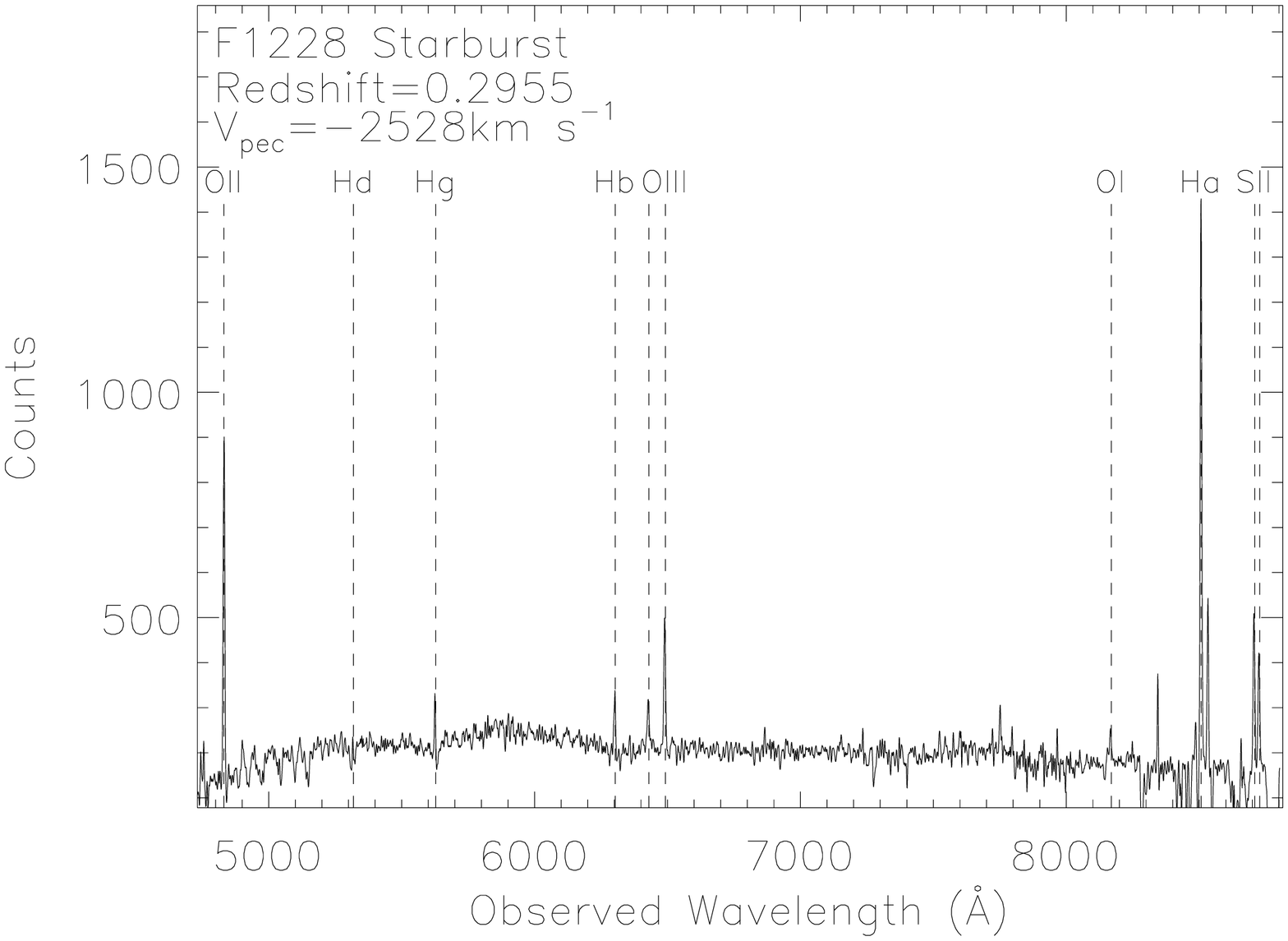}}
\caption{{\it Top panel:} Close up views of the jellyfish galaxies highlighted in Figure~\ref{hst_plus_chan}. The green contours show the surface brightness at $\sim 1 \sigma$ above the background for an image generated by co-adding the F435W, F606W and F814W images using the SWarp tool \citep{bertin2002}. The white circle shows the AAOmega fiber aperture size. {\it Bottom panel:} AAOmega spectra (where available) for the jellyfish galaxies. We note that the broad feature at $\sim 8650$\AA\, seen in the F0237 and F1228 spectra is due to sky subtraction residuals.}
\label{rgbimages}
\end{figure*}
\end{turnpage}

Close inspection of the HST images revealed four galaxies (marked by red boxes in Figure~\ref{hst_plus_chan}) with distinct trails of extremely blue knots and filaments, most conspicuous in the bluest F435W band and having magnitudes $24.7 < {\rm F435W} < 28.5$. This band corresponds to the Sloan $u$-band in the cluster rest frame, which is known to be sensitive to light from young, hot OB stars, hence active regions of star formation \citep{hopkins2003}. The top panels of Figure~\ref{rgbimages} show images of these four galaxies, revealing  their ``jellyfish'' morphology \citepalias[nomenclature of][]{smith2010}, which is due to trails and filaments of bright star-forming regions. This interpretation is consistent with star-forming signatures in the optical spectra discussed below for three of the four galaxies. All 3 galaxies are spectroscopically confirmed cluster members \citepalias{owers2011}.

Galaxy F0083 is a luminous ($\sim 3L^*_R$), nearly face-on late-type spiral with numerous blue star-forming regions in its disk and an unresolved bright nucleus. The disk structure is irregular with an asymmetry on the eastern side connecting to a faint tidal feature which connects to a nearby faint galaxy (F606W mag $\sim 22.6$), indicating an interaction has taken place. There is a bright blue rim $\sim 4$\,kpc north of the galaxy center. A trail of blue knots and filaments extends $\sim 35$\,kpc from the galaxy center to the southwest -- these knots and filaments are not associated with the disk. The spectrum (Figure~\ref{rgbimages}) reveals Balmer lines with narrow and broad components which are well fitted by two Gaussians having rest-frame velocity dispersions of $1078\pm13$\kms\ ($1155\pm40$\kms) and $124\pm2$\kms\ ($133\pm4$\kms), respectively, for \ha\, (\hb). The OIII doublet is also well fitted by broad and narrow components with dispersions $336\pm5$\kms\ and $93\pm1$\kms, respectively. The combination of broad and narrow lines indicate that F0083 hosts a Seyfert 1 nucleus. However, the ratios of the equivalent widths (EW) of the narrow components of log([NII]/\ha)=-0.51 and log([OIII]/\hb)=0.67 lie close to the AGN/star-forming boundary, indicating that a fraction of the emission may be due to star formation \citep{baldwin1981}. A bright X-ray point source is associated with this galaxy (Figure~\ref{hst_plus_chan}).

F0237 \citep[originally CN104;][]{couch1987} is a major merger involving two $\sim 0.7L^*_R$ late-type galaxies \citep[classified by][as Sbc and Sab]{couch1998} the centers of which are separated by $\sim 5$kpc. The multi-color image (Figure~\ref{rgbimages}) reveals a faint tidal feature extending $\sim 30$\,kpc to the southeast, and a trail of $\sim6$ blue knots extending $\sim21$\,kpc to the southwest. Unlike F0083, there are no blue knots within the galaxy disks. The AAOmega spectrum confirms \citeauthor{couch1987}' poststarburst classification---there is strong Balmer absorption (EW ${\rm H}\delta \sim 6$) with no measurable OII or \ha\ emission. The fiber aperture covers only a portion of F0237, leaving the extent of its poststarburst emission undetermined. However, the lack of blue star-forming regions within the system indicates a dearth of ongoing star formation.

The faint ($\sim 0.3L^*_R$), blue galaxy F1228 is the western-most of the highlighted galaxies in Figure~\ref{hst_plus_chan}. It hosts the least spectacular trail of blue knots, the most distant of which is $\sim 11$\,kpc from F1228's main component. The morphology is similar to the ``tadpole'' galaxies seen in the Hubble Ultra-Deep Field \citep{straughn2006,elmegreen2010} with a head that contains a number of bright knots of star formation, a blue ridge on its western side, and a tail pointing to the SE containing the blue knots. The spectrum (Figure~\ref{rgbimages}) reveals strong \ha\ emission with EW $\sim -52\AA$. The emission line EW ratios of [NII]/\ha=-0.64 and [OIII]/\hb=-0.04 are consistent with a star-forming origin \citep{baldwin1981}. 

The jellyfish galaxy closest to the Bullet-like subcluster (right-most panel of Figure~\ref{rgbimages}) has a disk-like morphology, with no discernible bulge or spiral arms. The trail of extremely blue knots points to the NW and extends to a radius of $\sim 26$\,kpc. This galaxy is below the brightness limit used in \citetalias{owers2011} and thus does not have a spectrum or, therefore, a redshift.

\subsection{Colors}\label{colours}

\begin{figure*}
{\includegraphics[angle=90,width=0.9\textwidth]{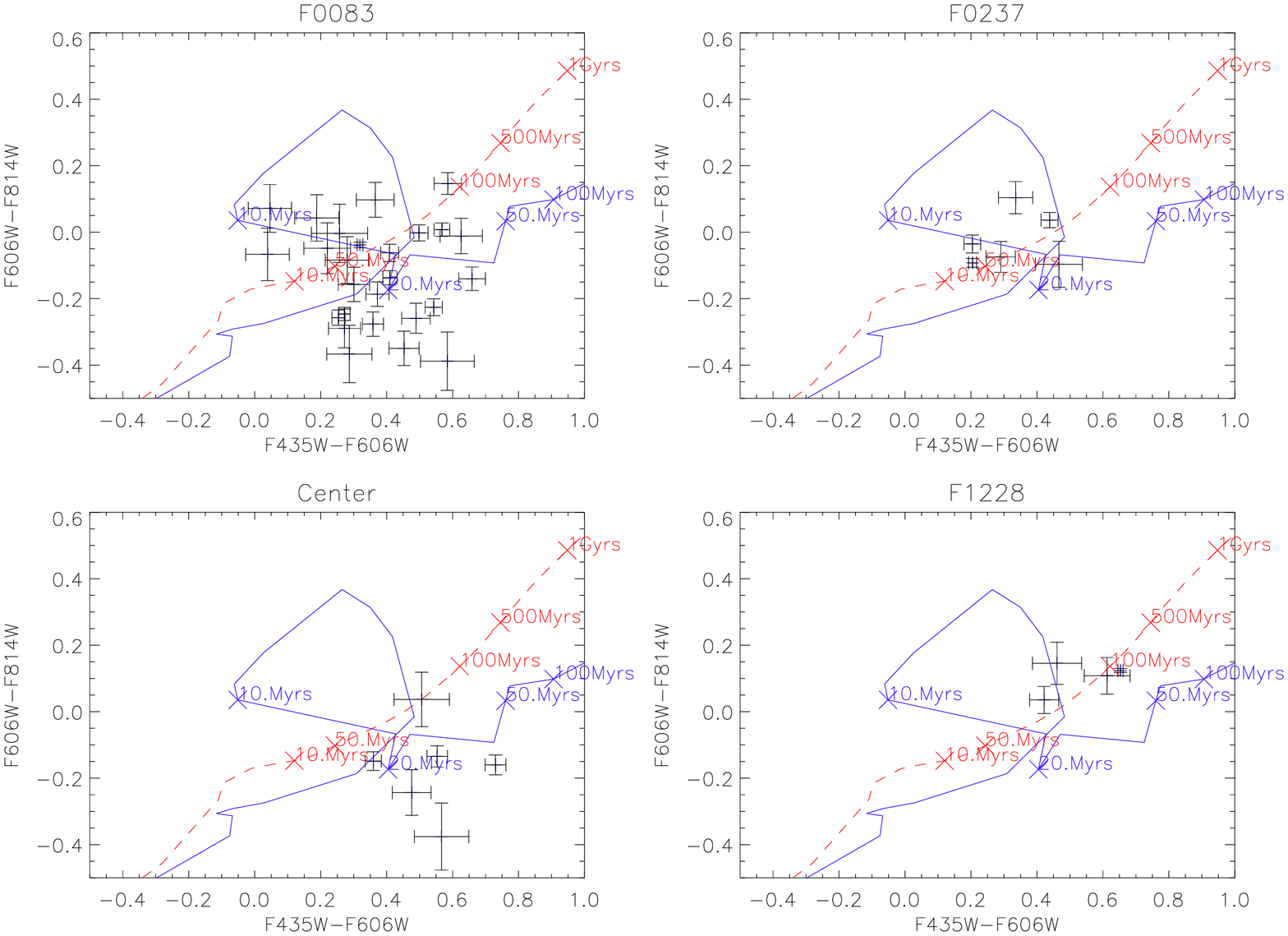}}
\caption{Comparison of the colors of the knots and filaments in the jellyfish tails (black symbols with error bars) to those predicted by models of stellar populations produced by single burst (solid blue line) and exponentially declining (red dashed line) star formation histories.}
\label{fig:colours}
\end{figure*}

We estimate the ages of the knots and filaments by comparing their colors to those predicted by the solar metallicity models of \citet{maraston2005}. The models used correspond to the stellar populations produced by two extremal star formation histories -- a single burst, and an exponentially decaying star formation with a decay timescale of 20\,Gyrs. The colors of the knots and filaments and the color evolution tracks derived from the models are shown in Figure~\ref{fig:colours}. There are two caveats. First, the contribution of emission lines is ignored, although this is exepected to be minimal within the broad bands of the filters \citep[e.g., see Figure 9][]{cortese2007}. Second, we do not account for reddening. However, dust obscuration tends to make stellar populations appear older, thus our age estimates are upper limits. From Figure~\ref{fig:colours} it can be seen that in the majority of cases the knots/filaments are bluer than predicted by the 100\,Myr populations, thus we take this to be the upper limit on the times since the onset of star formation.

\section{Discussion and Interpretation}

We have presented HST observations which reveal four ``jellyfish'' galaxies in the merging cluster A2744. Low redshift jellyfish analogues are observed in nearby clusters \citep{sun2007,yoshida2008, smith2010, hester2010}. However, at redshifts comparable to A2744, only three have been observed, all in separate clusters \citep[][]{owen2006, cortese2007}. Indeed, \cite{cortese2007} found only two examples in a survey of 13 intermediate redshift clusters. The rarity of jellyfish galaxies in intermediate redshift clusters emphasizes the significance of the observations presented here; we observe four such systems within A2744. While the statistics at intermediate redshifts are sparse, \citetalias{smith2010} compiled a sample of 13 Coma cluster galaxies which harbor ultraviolet asymmetries/tails. The orientation of the tails,  which generally point away from the cluster center, suggested that these galaxies formed from an infalling population experiencing the cluster environment for the first time \citepalias{smith2010}. However, there are two key distinctions to be made when comparing the low redshift jellyfish population with that of A2744. First, the knots in A2744 are bright ($-13 \lesssim M_{I} \lesssim -17$) when compared to prominent examples in Coma \citep[GMP4060/RB199, $M_{I} > -13$ S10;][]{yoshida2012} and Virgo \citep[IC3418, $M_{I} > -11$][]{hester2010,fumagalli2011} while only ESO 137-001 in the merging cluster A3627 has knots with comparable brightness \citep{sun2007,woudt2008}. Thus, the bright knots seen in A2744 are rare in low redshift clusters. Second, the orientation of the tails in A2744 are less well ordered than those seen in Coma. This does not preclude the existence of a faint, Coma-like infalling population of jellyfish in A2744, which may be revealed by deeper observations, but indicates that the picture outlined by \citetalias{smith2010} is unlikely in this case for the jellyfish observed in A2744. Thus, there appears to be a relatively larger number of jellyfish in A2744 compared to other intermediate redshift clusters and these jellyfish appear to be a different, brighter version of their low redshift counterparts. 

Is the major merger in A2744 driving the formation of an excess of these bright jellyfish galaxies? There is a strong spatial correlation between the jellyfish galaxies and features associated with the high speed Bullet-like subcluster \citepalias[Figure~\ref{hst_plus_chan} and][]{owers2011}: the proximity of (i) F0083 and F0237 to the portion of the Bullet-driven shock front revealed as an edge in the \chan\ observations, and (ii) the central jellyfish to the X-ray peak associated with the remnant gas core of the Bullet-like subcluster. While we cannot know the exact 3D locations of the jellyfish galaxies with respect to the ICM structures, the small projected distances suggest that the jellyfish galaxies may have recently been overrun by the shock front and/or the Bullet-like subcluster gas. This indicates that a mechanism related to an interaction with these ICM features may be responsible for either the stripping of the gas leading to the tails or the triggering of the star formation in the tails, or both. This assertion is supported by the young ages of the stellar populations in the knots and filaments, which suggest that the star formation was triggered $\lesssim 100\,{\rm Myr}$ ago (Section~\ref{colours}). On these timescales, a galaxy with velocity $\sim 1000$\kms\ travels $\lesssim 100$\,kpc, so we would expect there to still be a strong spatial coincidence between the jellyfish and the putative ICM features responsible for triggering the star formation. Furthermore, consideration of the peculiar velocity of the Bullet-like subcluster \citepalias[$v_{pec} \simeq 2500$\kms;][]{owers2011} and of the three jellyfish galaxies with measured redshifts ($v_{pec} = -729, -2277\, {\rm and} -2528$\kms; lower panels, Figure~\ref{rgbimages}) indicates that the jellyfish galaxies are not members of the Bullet-like subcluster, and that if they have interacted with the shock or the Bullet-like subcluster's ICM, then the relative velocity of the interaction was high -- of the order of the merger velocity $\sim 4750$\kms. Similarly, \citet{owen2006} suggested that the jellyfish-like galaxy C153 in A2125 may be a result of enhanced ram pressure stripping caused by a high velocity encounter with the ICM due to a cluster merger, while \citetalias{smith2010} find hints that some of their jellyfish galaxies are associated with merger-related enhancements in the ICM density.

The proximity of the jellyfish to merger-related ICM features in this high speed merger suggests that the merger is responsible for the increased fraction and more extreme nature of the jellyfish in A2744. A high-speed, head-on merger creates much greater ram pressure and a powerful shock which can significantly enhance galaxy-ICM processes when compared with those felt by a galaxy falling into a relaxed cluster. Enhanced ram pressure may strip gas more efficiently leading to a higher incidence of gaseous tails, and therefore a higher likelihood of forming the observed jellyfish phenomena. This is consistent with the simulations of \citet{domainko2006}, who find the mass lost from a galaxy due to ram pressure during a major merger is substantially increased, and also with \citet{vollmer2006} who suggest the enhanced ram pressure stripping due to an interaction with the infalling M49 group explains the strong ram pressure stripping observed in NGC~4522. Furthermore, the increased ICM pressure due to the merger shock may promote star formation in the tails and may also drive higher star formation rates, hence higher surface brightness features, compared with jellyfish galaxies in lower pressure environments. The shock has a Mach number $M \simeq 3$ \citepalias{owers2011} meaning the pressure jumps by a factor of $\sim 11$ with respect to the pressure of the surrounding unshocked ICM, $P_{\rm ICM}/{\rm k_B} \sim 10^5 {\rm K/cm^3}$. Thus, the static pressure due to the shock is $P_{\rm Shock}/{\rm k_B} \sim 10^6 {\rm K/cm^3}$ which is an order of magnitude higher than the threshold pressure required to trigger the collapse of GMCs leading to star formation \citep{elmegreen1997,bekki2010}. Furthermore, there is a rapid increase in the ram-pressure the galaxy feels as it is overrun by the shock due to the high relative velocity and the factor of 3 increase in ICM density for a Mach~3 shock. If, like the Bullet cluster \citep{Markevitch2007}, the shock front is collisionless, it would be more abrupt than a collision-dominated shock (relaxation time $\sim 10$\,Myr), so the interaction is likely too fast ($\sim 10$\,Myr) for the shock's ram pressure to be responsible for stripping the gas leading to the observed tails. This is supported by the southwesterly orientation of the trails of material seen in F0083 and F0237. If ram pressure from the shock was responsible, the tails should point southeast toward the near part of the shock front. This high pressure shock-triggering scenario is consistent with the simulations of \citet{tonnesen2012} and \citet{kapferer2009} who found that the star formation rates in ram-pressure stripped gaseous tails is largely driven by the ICM pressure. However, simulations specific to the scenario outlined above are required for confirmation.

On the question of whether the increased ram pressure which occurs in mergers enhances or suppresses the star formation in cluster members \citep{fujita1999,bekki2003,kronberger2008,bekki2010}, the observations presented here show that both effects may be in play. Considering F0083, a number of blue knots are seen across the galaxy indicating that there is disk-wide star formation. There is evidence for an interaction with a smaller galaxy (Figure~\ref{rgbimages}), which may have driven gas towards the galaxy center, triggering the AGN activity. However, such a minor interaction is unlikely to trigger disk-wide star formation. Thus, we suggest that the high ICM pressure has compressed the GMCs and triggered disk-wide star formation \citep{bekki2003}. The asymmetric distribution in the star formation---particularly the arc-shaped star-forming region $\sim 4$\,kpc to the north of the galaxy center on the side opposing the tail of star-forming regions---is consistent with that expected due to the compression from ram pressure as a galaxy moves edge-on through the ICM \citep{kronberger2008}. Conversely, the spectrum of F0237 (Figure~\ref{rgbimages}) shows strong Balmer absorption and an absence of emission lines indicating a starburst has recently ($<1$\,Gyr ago) been abruptly truncated. Here, the major merger of two spiral galaxies is the most likely trigger for the initial starburst \citep{mihos1996,couch1998, bekki2001a,owers2007}. However, in the absence of external mechanisms the poststarburst phase is expected to occur at the later stages of the major merger after the galaxies have coalesced \citep{bekki2001b,blake2004, snyder2011}. The premature truncation of the starburst may be due to the enhanced ram pressure felt by a galaxy during a cluster merger, which strips gas more rapidly. Furthermore, the high pressure cluster merger environment may increase the star formation leading to a more rapid consumption of the gas. The difference in the star-forming properties of F0083 and F0237 may be attributed to the different mass and dynamical states of the two systems. Given its luminosity, F0083 is likely more massive than F0237, meaning it is less susceptible to complete stripping of gas due to ram pressure. Further, strong tidal forces due to the major merger in F0237 may unbind gas, making it more susceptible to complete ram pressure stripping \citep{vollmer2003,kapferer2008} thereby halting star formation in the disks. 

\section{Conclusions}

We have presented HST observations of four rare ``jellyfish'' galaxies in the merging cluster A2744. The knots and filaments are brighter than the majority of jellyfish analogues found in local clusters, indicating more vigorous star formation is occurring compared to that found in tails of low redshift jellyfish. Intriguingly, three of these jellyfish galaxies lie in close proximity to merger-related features in the ICM. This leads us to propose that the rise of the jellyfish in A2744 is due to the effects of the high pressure of the ICM during the ongoing cluster merger. In particular, we suggest that the star formation occurring in the tails of stripped gas has been triggered by the rapid, dramatic increase in pressure during an interaction with the shock. These observations support the hypothesis that galaxies undergo accelerated evolution in their star-forming properties during the violent core passage phase of a cluster merger.

We thank the referee, Curtis Struck, for his helpful comments. M.S.O. and W.J.C acknowledge the financial support of the Australian Research Council. P.E.J.N. was partly supported by NASA grant NAS8-03060. 

{\it Facilities:} \facility{CXO (ACIS)}, \facility{AAT (AAOmega)}, \facility{HST (ACS)}


\end{document}